# The Traveling Salesman Problem: A Linear Programming Formulation

MOUSTAPHA DIABY
Operations and Information Management
University of Connecticut
Storrs, CT 06268
USA
moustapha.diaby@business.uconn.edu

*Abstract:* - In this paper, we present a polynomial-sized linear programming formulation of the *Traveling Salesman Problem* (TSP). The proposed linear program is a network flow-based model. Numerical implementation issues and results are discussed.

## 1 Introduction

The Traveling Salesman Problem (TSP) is the problem of finding a least-cost sequence in which to visit a set of cities, starting and ending at the same city, and in such a way that each city is visited exactly once. This problem has received a tremendous amount of attention over the years due in part to its wide applicability in practice (see Lawler et *al.* [1985] among others, for examples). Also, since its seminal formulation as a mathematical programming problem in the 1950's (Dantzig, Fulkerson, and Johnson [1954]), the problem has been at the core of most of the developments in the area of Combinatorial Optimization (see Nemhauser and Wolsey [1988], among others). A key issue has been the question of whether there exists a polynomial-time algorithm for solving the problem (see Garey and Johnson [1979]).

In this paper, we present a polynomial-sized linear programming formulation of the Traveling Salesman Problem (TSP). The proposed linear program is a network flow-based model. Numerical implementation issues and results are discussed.

The plan of the paper is as follows. The proposed linear programming formulation is developed in section 2. Numerical implementation and computational results are discussed in section 3. Conclusions are discussed in section 4.

## 2 Problem Formulation

Different classical formulations of the TSP are analyzed and compared in Padberg and Sung [1991]. The approach used in this paper is different from that of any of the existing models that we know of. In this section, we first present a nonlinear integer programming (NIP) formulation of the TSP. Then, we develop an integer linear programming (ILP) reformulation of this NIP model using a network flow modeling framework. Finally, we show that the linear programming (LP) relaxation of our ILP reformulation has extreme points that correspond to TSP tours respectively.

### 2.1 NIP Model

Consider the TSP defined on $n$ nodes belonging to the set $N = \{1, 2, \ldots, n\}$, with arc set $E = N^2$, and travel costs $t_{ij}$ ($(i,j) \in E$; $t_{ii} = \infty$, $\forall\ i \in N$) associated with the arcs. Assume, without loss of generality, that city 1 is the starting point and the ending point of travel. Denote the set of the remaining cities as $M = N \setminus \{1\}$. Define $S = N \setminus \{n\}$ as the index set for the stage of travel corresponding to the order of visit of the cities in M. Let $R \equiv S \setminus \{n\text{-}1\}$.

Let $u_{is}$ ($i \in M$, $s \in S$) be a 0/1 binary variable that takes on the value "1" if city $i \in M$ is visited at stage $s \in S$. Then, in order to properly account the TSP travel costs, consecutive travel stages must be considered jointly. Hence, re-define the travel costs as:

$$c_{isj} = \begin{cases} t_{ij} + t_{1,i},\ s = 1,\ (i,j) \in M^2; \\ t_{ij},\ s \in R \setminus \{1, n-2\}, (i,j) \in M^2; \\ t_{ij} + t_{j,1},\ s = n-2,\ (i,j) \in M^2. \end{cases} \qquad (2.1)$$

Then, the cost incurred if city $i \in M$ is visited at stage $s \in R$ followed by city $j \in M$ at stage $(s+1)$ can be expressed as $c_{isj}u_{is}u_{j,s+1}$ ($(i, j) \in M^2$, $s \in R$). For example, $c_{235}u_{23}u_{54}$ would represent the cost function associated with the situation where cities 2 and 5 are the 3rd and 4th cities to be visited (after city 1), respectively.

Note that from expression 2.1 above, $c_{i,1,j}u_{i,1}u_{j,2}$ and $c_{i,n-2,j}u_{i,n-2}u_{j,n-1}$ correctly model the costs of the travels $1 \to i \to j$ and $i \to j \to 1$, respectively. Hence, the TSP can be formulated as the following nonlinear bipartite matching problem.

**Problem TSP:**

Minimize

$$Z_{TSP}(u) = \sum_{s \in R} \sum_{i \in M} \sum_{j \in (M \setminus \{i\})} c_{isj} u_{is} u_{j,s+1} \quad (2.2)$$

Subject to:

$$\sum_{i \in M} u_{is} = 1 \quad s \in S \quad (2.3)$$

$$\sum_{s \in S} u_{is} = 1 \quad i \in M \quad (2.4)$$

$$u_{is} \in \{0, 1\} \quad i \in M;\ s \in S \quad (2.5)$$

The objective function 2.2 aims to minimize the total cost of all travels. Constraints 2.3 stipulate (in light of the binary requirements constraints 2.5) that only one city can be visited from city 1 and that only one city is visited at each stage of travel. Constraints 2.4 on the other hand ensure (in light of the binary requirements 2.5) that a given city is visited at exactly one stage of travel. The quadratic objective function terms (i.e., the $c_{isj}u_{is}u_{j,s+1}$'s) ensure (in light of the binary requirements constraints 2.5) that a travel cost is incurred from city i to city j iff those two cities are visited at consecutive stages of travel with i preceding j, as discussed above. Hence, *Problem TSP* accurately models the TSP.

## 2.2 ILP Model

Note that the polytope associated with *Problem TSP* is the standard assignment polytope (see Bazaraa, Jarvis, and Sherali [1990; pp. 499-513]), and that there is a one-to-one correspondence between TSP tours and extreme points of this polytope. Our modeling consists essentially of *lifting* this polytope in higher dimension in such a way that the quadratic cost function of *Problem TSP* is correctly captured using a linear function. To do this, we use the framework of the graph G = (V, A) illustrated in Figure 2.1, where the nodes in V correspond to (city, travel stage) pairs (i, s) ∈ (M, S), and the arcs correspond to binary variables $x_{irj} = u_{ir}u_{j,r+1}$ ((i, j) ∈ (M, M\{i}); r ∈ R). Clearly, there is a one-to-one correspondence between the perfect bipartite matching solutions of *Problem TSP* (and therefore, TSP tours) and paths in this graph that simultaneously span the set of stages, S, and the set of cities, M. For simplicity of exposition we refer to such paths as "*city and stage spanning*" ("*c.a.s.s.*") paths. Also, we refer to the set of all the nodes of the graph that have a given city index in common as a "level" of the graph, and to the set of all the nodes of the graph that have a given travel stage index in common as a "stage" of the graph.

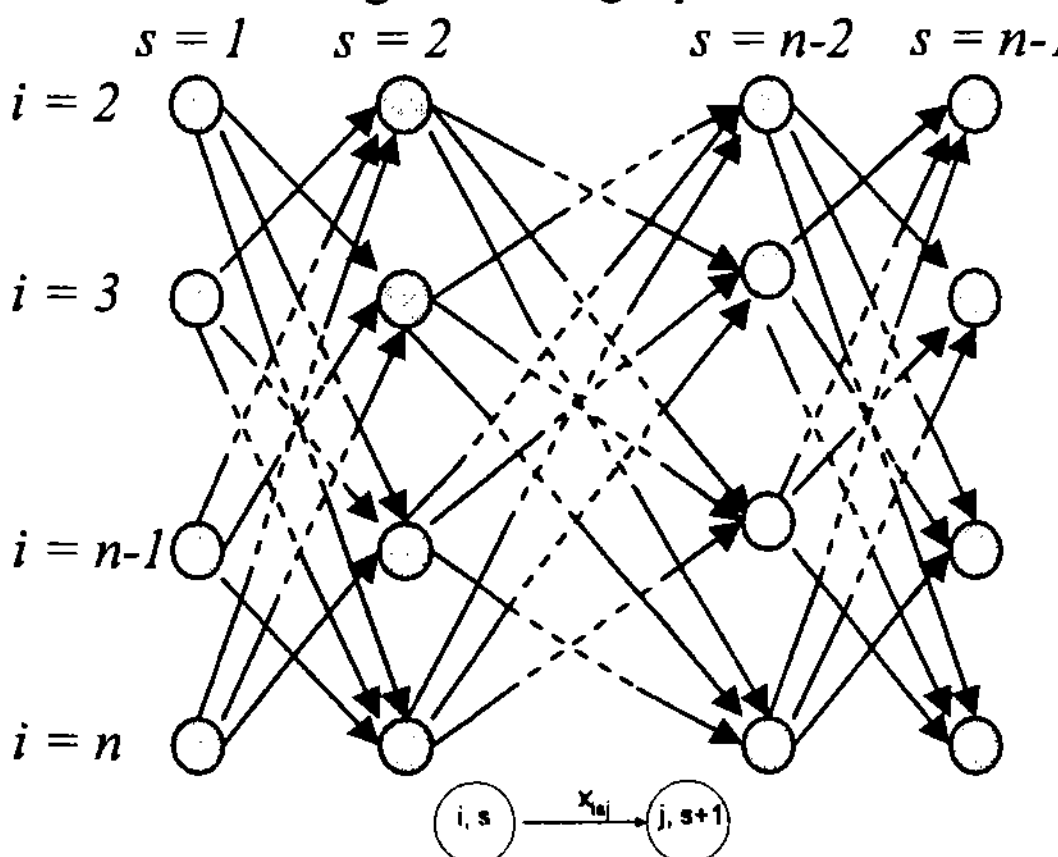

Fig. 2.1: *Illustration of Graph G*

The idea of our approach to reformulating *Problem TSP* is to develop constraints that "force" flow in Graph G to propagate along *c.a.s.s. paths* of the graph only. Hence, we do not deal directly with the TSP polytope *per se* (see Grötschel and Padberg [1985, pp. 256-261]) in this paper. More specifically, our approach in the paper consists of developing a reformulation of the polytope described by constraints 2.3 – 2.5 (i.e., the standard assignment polytope) using variables that are functions of the flow variables associated with the arcs of Graph G. The correspondence between vertices of our model and TSP tours is achieved through the association of costs to the vertices of the model, much in the same way as is done in *Problem TSP*. Therefore, developments that are concerned with descriptions of the TSP polytope specifically (see Padberg and Grötschel [1985], or Yannakakis [1991] for example) are not applicable in the context of this paper.

For (i, j, u, v, k, t) ∈ $M^6$, (p, r, s) ∈ $R^3$ such that r < p < s, let $z_{irjupvkst}$ be a 0/1 binary variable that takes on the value "1" if and only if the flow on arc (i, r, j) of Graph G subsequently flows on arcs (u, p, v) and (k, s, t), respectively. Similarly, for (i, j, k, t) ∈ $M^4$, (s, r) ∈ $R^2$ such that r < s, let $y_{irjkst}$ be a binary variable that indicates whether the flow on arc (i, r, j) subsequently flows on arc (k, s, t) ($y_{irjkst}$ = 1) or not ($y_{irjkst}$ = 0). Finally, denote by $y_{irjirj}$ the binary variable that indicates whether there is flow on arc (i, r, j) or not. Given an instance of (y, **z**), we use the term "flow layer" to refer to the sub-graph of G induced by the arc (i, r, j) corresponding to a given positive $y_{irjirj}$ along with the arcs (k, s, t) ($s \in R$, $s > r$) corresponding to the corresponding $y_{irjkst}$'s that are positive. Hence, the flow on arc (i, r, j) also flows on arc (k, s, t) (for a given $s > r$) iff arc (k, s, t) belongs to the *flow layer* originating from arc (i, r, j). Also, we say that flow on a given arc (i, r, j) of Graph G "visits" a given *level* of the graph, *level* t, if $\sum_{s \in R;\, s<r} \sum_{k \in M \setminus \{i,j,t\}} y_{tskirj} + \sum_{s \in R;\, s>r} \sum_{k \in M \setminus \{i,j,t\}} y_{irjkst} > 0$.

Logical constraints of our model are that: 1) flow must be conserved; 2) *flow layers* must be consistent with one another; and, 3) flow must be connected. For $(i, r, j) \in A$ such that $y_{irjirj} > 0$ in a given instance of (**y**, **z**), and $s > r$ ($s \in R$), define $F_s(i,\ r,\ j) \equiv \{(k,t) \in M^2 \mid y_{kstirj} > 0\}$. Then, by "consistency of flow layers" we are referring to the condition that the *flow layer* originating from arc (i, r, j) must be a sub-graph of the union of the *flow layers* originating from the arcs comprising each of the $F_s(i,\ r,\ j)$'s, respectively. In addition to the logical constraints, the bipartite matching constraints 2.3 and 2.4 of *Problem TSP* must be respectively enforced. These ideas are developed in the following.

1) *Flow Conservations.* Any flow through Graph G must be initiated at *stage 1*. Also, for $(i, j) \in M^2$, $r \in R$, $r \geq 2$, the flow on arc (i, r, j) must be equal to the sum of the flows from *stage 1* that propagate onto arc (i, r, j):

$$\sum_{i \in M} \sum_{j \in M} y_{i,1,j,i,1,j} = 1 \qquad (2.6)$$

$$y_{irjirj} - \sum_{u \in M} \sum_{v \in M} y_{u,1,virj} = 0;$$
$$i, j \in M;\ r \in R, r \geq 2 \qquad (2.7)$$

2) *Consistency of "Flow Layers".* For $p, s \in R$ ($1 < p < s$) and $(u, v, k, t) \in M^4$, flow on (u, p, v) subsequently flows onto (k, s, t) iff for each $r < p$ ($r \in R$) there exists $(i, j) \in M^2$ such that flow from (i, r, j) propagates onto (k, s, t) via (u, p, v). This results in the following three types of constraints:

i) *Layering Constraints A*

$$y_{irjupv} - \sum_{k \in M} \sum_{t \in M} z_{irjupvkst} = 0;$$
$$i, j, u, v \in M;\ p, r, s \in R, 2 \leq p \leq n\text{-}3,$$
$$r \leq p\text{-}1,\ s \geq p\text{+}1 \qquad (2.8)$$

ii) *Layering Constraints B*

$$y_{irjkst} - \sum_{u \in M} \sum_{v \in M} z_{irjupvkst} = 0;$$
$$i, j, k, t \in M;\ p, r, s \in R, 2 \leq p \leq n\text{-}3;$$
$$r \leq p\text{-}1,\ s \geq p\text{+}1 \qquad (2.9)$$

iii) *Layering Constraints C*

$$y_{upvkst} - \sum_{i \in M} \sum_{j \in M} z_{irjupvkst} = 0;$$
$$u, v, k, t \in M;\ p, r, s \in R, 2 \leq p \leq n\text{-}3,$$
$$r \leq p\text{-}1,\ s \geq p\text{+}1 \qquad (2.10)$$

3) *Flow Connectivities.* All flows must propagate through the graph, on to stage n-1, in a connected manner. Each *flow layer* must be a connected graph, and must conserve flow:

$$\sum_{k \in M} y_{irjkst} - \sum_{k \in M} y_{irjt,s+1,k} = 0;\ i, j, t \in M;$$
$$r, s \in R, r \leq n\text{-}3, r \leq s \leq n\text{-}3 \qquad (2.11)$$

$$\sum_{v \in M} z_{vpuirjkst} - \sum_{v \in M} z_{u,p+1,virjkst} = 0;$$
$$i, j, k, t, u \in M;\ p, r, s \in R,$$
$$3 \leq r \leq n\text{-}3, s \geq r\text{+}1,\ p \leq r\text{-}2 \qquad (2.12)$$

$$\sum_{v \in M} z_{irjvpukst} - \sum_{v \in M} z_{irju,p+1,vkst} = 0;$$
$$i, j, k, t, u \in M;\ p, r, s \in R,$$
$$r \leq n\text{-}5, s \geq r\text{+}3,\ r\text{+}1 \leq p \leq s\text{-}2 \qquad (2.13)$$

$$\sum_{v \in M} z_{irjkstvpu} - \sum_{v \in M} z_{irjkstu,p+1,v} = 0;$$
$$i, j, k, t, u \in M;\ p, r, s \in R,$$
$$r \leq n\text{-}5, r\text{+}1 \leq s \leq n\text{-}4,\ s\text{+}1 \leq p \leq n\text{-}3 \qquad (2.14)$$

4) *"Visit" Requirements.* Flow within any *layer* must *visit* every *level* of Graph G:

$$y_{irjkst} - \sum_{p \in R;\, p \leq r-1} \sum_{v \in M} z_{upvirjkst} +$$
$$- \sum_{p \in (R \cap [r+1, s-2])} \sum_{v \in M} z_{irjvpukst} +$$
$$- \sum_{p \in R;\, p \geq s+1} \sum_{v \in M} z_{irjkstvpu} = 0;\ r, s \in R, s \geq r\text{+}1;$$
$$i, j, k, t \in M;\ u \in M \setminus \{i, j, k, t\} \qquad (2.15)$$

5) *"Visit" Restrictions.* Flow must be connected with respect to the stages of Graph G. There can be no flow between nodes belonging to the same *level* of the graph; No *level* of the graph can be *visited* at more than one *stage*, and vice versa:

$$\sum_{(k,t) \in M^2 | (k,t) \neq (i,j)} y_{irjkrt} + \sum_{s \in R;\, s \geq r+1} \sum_{k \in M} y_{irjksi} +$$
$$+ \sum_{s \in R;\, s \geq r+1} \sum_{k \in M} y_{irjisk} + \sum_{s \in R;\, s \geq r} \sum_{k \in M} \sum_{t \in M} y_{irikst} +$$
$$+ \sum_{(k,t) \in (M \setminus \{j\}, M) | (k, r+1, t) \in A} y_{irjk,r+1,t} +$$

$$+ \sum_{s\in R; s\geq r+1} \sum_{k\in M} y_{irjksj} + \sum_{s\in R; s\geq r+2} \sum_{k\in M} y_{irjjsk} +$$
$$+ \sum_{s\in R; s\leq r} \sum_{k\in M} \sum_{t\in M} y_{kstjrj} = 0,$$
$$i, j \in M; r \in R \tag{2.16}$$

Note that constraints 2.3 of *Problem TSP* are enforced through the combination of the "Flow Connectivities" requirements 2.11 – 2.14 and the *'Visit' Restrictions* constraints 2.16, and that constraints 2.4 are enforced through the *'Visit' Requirements* constraints 2.15.

The complete statement of our integer (linear) programming model is as follows:

***Problem IP***:

Minimize

$$Z_{IP}(\mathbf{y}, \mathbf{z}) = \sum_{r\in R} \sum_{i\in M} \sum_{j\in M} c_{irj} y_{irjirj}$$

Subject to:

Constraints 2.6 – 2.16

$y_{irjkst}, z_{irjupvkst} \in \{0, 1\}$ $i, j, k, t, u, v \in M;$

$p, r, s \in R$

The following theorem formally establishes the equivalence between *Problem IP* and *Problem TSP*.

**Theorem 1**

*Problem IP* and *Problem TSP* are equivalent.

*Proof:*

i) For a feasible solution to *Problem TSP*, $\mathbf{u} = (u_{is})$, let $(\mathbf{y}(\mathbf{u}), \mathbf{z}(\mathbf{u}))$ be a vector with components specified as follows:

$$\begin{cases} (\mathbf{y}(\mathbf{u}))_{irjkst} = u_{ir} u_{j,r+1} u_{ks} u_{t,s+1}; \\ \quad i, j, k, t \in M; r, s \in R, s \geq r \\ (\mathbf{z}(\mathbf{u}))_{irjapbkst} = u_{ir} u_{j,r+1} u_{ap} u_{b,p+1} u_{ks} u_{t,s+1}; \\ \quad a, b, i, j, k, t \in M; \quad p, r, s \in R, \ r < p < s \end{cases}$$

It is easy to verify that $(\mathbf{y}(\mathbf{u}), \mathbf{z}(\mathbf{u}))$ satisfies each of the constraints of *Problem IP*.

ii) Let $(\mathbf{y}, \mathbf{z}) = (y_{irjkst}, z_{abirjkst})$ be a feasible solution to *Problem IP*. Because of constraints 2.6, 2.7, 2.11, and the binary requirements on the variables, $(\mathbf{y}, \mathbf{z})$ must be such that there exists a set of city indices $\{i_1, i_2, \cdots, i_{n-1}\}$ with:

$$y_{i_r, r, i_{r+1}, i_r, r, i_{r+1}} = 1 \quad \forall r \in R$$

Because of constraints 2.8 - 2.10, and the binary requirements, we must also have:

$$y_{arbcsd} = \begin{cases} 1 & \text{for } (a, b, c, d) = (i_r, i_{r+1}, i_s, i_{s+1}), \\ 0 & \text{otherwise} \end{cases}$$

$\forall (r, s) \in R^2$ with $r < s$; and

$$z_{arbcpdesf} = \begin{cases} 1 & \text{for } (a, b, c, d, e, f) = \\ & \quad (i_r, i_{r+1}, i_p, i_{p+1}, i_s, i_{s+1}); \\ 0 & \text{otherwise} \end{cases}$$

$\forall (r, p, s) \in R^3$ with $r < p < s$

Hence, by constraints 2.16, the $i_s$'s must be such that:

$i_r \neq i_s$ for all $(r, s) \in R^2$ such that $s \neq r$.

Hence, a unique feasible solution to *Problem TSP* is obtained from $(\mathbf{y}, \mathbf{z})$ by setting:

$$u_{jr} = \begin{cases} 1 & \text{if } j = i_r \\ 0 & \text{otherwise} \end{cases} \quad \forall j \in M, r \in S$$

iii) Clearly, from i) and ii) above, *Problem IP* and *Problem TSP* have equivalent feasible sets. The theorem follows from this and the fact that the two problems also have equivalent objective functions.

Q.E.D.

Hence, each feasible solution to *Problem IP* corresponds to a TSP tour, and conversely. Let $\varphi(\ell) = \langle 1, \ell_1, \cdots, \ell_{n-1}, 1 \rangle$ denote the ordered set of city indices visited along a given TSP tour, Tour $\ell$ (i.e., with $\ell_t$ as the index of the city visited at stage t according to Tour $\ell$). In the remainder of this paper, we will use the term "feasible solution corresponding to (Given) Tour $\ell$" to refer to the vector $(\mathbf{y}(\varphi(\ell)), \mathbf{z}(\varphi(\ell)))$ obtained as follows:

$$(\mathbf{y}(\varphi(\ell)))_{arbcsd} = \begin{cases} 1 & \text{for } r, s \in R, \ s \geq r, \\ & \quad (a, b, c, d) = (\ell_r, \ell_{r+1}, \ell_s, \ell_{s+1}); \\ 0 & \text{otherwise} \end{cases}$$

$$(\mathbf{z}(\varphi(\ell)))_{apbcrdesf} = \begin{cases} 1 & \text{for } p, r, s \in R, \ \ s > r > p, \\ & \quad (a, b, c, d, e, f) = \\ & \quad = (\ell_p, \ell_{p+1}, \ell_r, \ell_{r+1}, \ell_s, \ell_{s+1}); \\ 0 & \text{otherwise} \end{cases}$$

Our proposed linear programming model will now be developed.

## 2.3 LP Model

Our basic linear programming model consists of the linear programming relaxation of *Problem IP*. This problem can be stated as follows:

***Problem LP:***

Minimize

$$Z_{LP}(\mathbf{y}, \mathbf{z}) = \sum_{i\in M} \sum_{r\in R} \sum_{j\in M} c_{irj} y_{irjirj} \qquad (2.17)$$

Subject to:

Constraints 2.6 – 2.16

$$y_{irjkst}, z_{upvirjkst} \in [0, 1]; \quad u, v, i, j, k, t \in M, \quad p, r, s \in R \qquad (2.18)$$

In the remainder of this section, we establish the equivalence between *Problem LP* and *Problem IP*. We begin with the following result.

**Lemma 1**

The following constraints are valid for *Problem LP*:

i) $y_{irjirj} - \sum_{k\in M} \sum_{t\in M} y_{irjkst} = 0;$

$\quad i, j \in M; \ r, s \in R, \ s \geq r+1$

ii) $y_{irjirj} - \sum_{k\in M} \sum_{t\in M} \sum_{a\in M} \sum_{c\in M} z_{irjkstabc} = 0;$

$\quad i, j \in M; \ r, s, b \in R, \ r < s < b$

*Proof:*

i) $y_{irjirj} = \sum_{u\in M} \sum_{v\in M} y_{u,1,virj}$;

$\quad i, j \in M; \ r \in R\backslash\{1\}$ (Using 2.7)

$\quad = \sum_{u\in M} \sum_{v\in M} \sum_{k\in M} \sum_{t\in M} z_{u,1,virjkst}$

$\quad i, j \in M; \ r \in R, \ 1 < r < s$ (Using 2.8)

$\quad = \sum_{k\in M} \sum_{t\in M} \sum_{u\in M} \sum_{v\in M} z_{u,1,virjkst}$

$\quad i, j \in M; r, s \in R, 1 < r < s$ (Re-arranging)

$\quad = \sum_{k\in M} \sum_{t\in M} y_{irjkst}$

$\quad i, j \in M; \ r, s \in R, \ 1 < r < s$ (Using 2.10)

Combining the above with constraints 2.11 (for r = 1), we have:

$y_{irjirj} = \sum_{k\in M} \sum_{t\in M} y_{irjkst}$; $i, j \in F; \ r, s \in R, \ s \geq r+1$

ii) Condition ii) follows directly from the combination of Lemma 1-i) and constraints 2.8.

Q.E.D.

For a feasible solution $(\mathbf{y}, \mathbf{z}) = (y_{irjkst}, z_{upvirjkst})$ to *Problem LP*, let *G*(**y**, **z**) = (V(**y**, **z**), A(**y**, **z**)) be the sub-graph of *G* induced by the arcs of G corresponding to the positive components of (**y**). For r ∈ R, define $W_r(\mathbf{y}, \mathbf{z}) \equiv \{(i, j) \in M^2 \mid \{(i, r, j) \in A(\mathbf{y}, \mathbf{z})\}$. Denote the arc corresponding to the $v^{th}$ element of $W_r(\mathbf{y}, \mathbf{z})$ $(v \in \{1, 2, \cdots, \chi_r(\mathbf{y}, \mathbf{z})\}; 1 \leq \chi_r(\mathbf{y}, \mathbf{z}) \leq (n-1)(n-2))$ as $a_{r,v}(\mathbf{y}, \mathbf{z}) = (i_{r,v}, r, j_{r,v})$. Then, $W_r(\mathbf{y}, \mathbf{z})$ can be alternatively represented as $X_r(\mathbf{y}, \mathbf{z}) = \{(i_{r,v}, r, j_{r,v}); \ v \in N_r(\mathbf{y}, \mathbf{z})\}$, where $N_r(\mathbf{y}, \mathbf{z}) = \{1, 2, \cdots, \chi_r(\mathbf{y}, \mathbf{z})\}$ is the index set for the arcs of *Graph* G(**y**, **z**) originating at *stage* r. For convenience, we will henceforth write $a_{r,v}(\mathbf{y}, \mathbf{z})$ simply as $a_{r,v}$. Furthermore, we will use a more compact indexing of the **y** and **z** variables where the set of indices "$i_{r,v}, r, j_{r,v}$" will be replaced with "($a_{r,v}$)", whenever convenient.

For (r, s) ∈ $R^2$ with s ≥ r+2, $\rho \in N_r(\mathbf{y}, \mathbf{z})$, and $\sigma \in N_s(\mathbf{y}, \mathbf{z})$ we refer to a set of arcs of G(**y**, **z**),

$$\begin{aligned} U_{(r,\rho),(s,\sigma),t}(\mathbf{y}, \mathbf{z}) \equiv \Big\{ & a_{r,v_{r,(r,\rho),(s,\sigma),t}}, a_{r+1,v_{r+1,(r,\rho),(s,\sigma),t}}, \\ & \cdots, a_{s,v_{s,(r,\rho),(s,\sigma),t}} \Big| \ v_{r,(r,\rho),(s,\sigma),t} = \rho; v_{s,(r,\rho),(s,\sigma),t} = \sigma; \\ & v_{p,(r,\rho),(s,\sigma),t} \in N_p(\mathbf{y}, \mathbf{z}), \forall p \in (R \cap [r+1, s-1]; \\ & i_{p,v_{p,(r,\rho),(s,\sigma),t}} = j_{p-1,v_{p-1,(r,\rho),(s,\sigma),t}}, \ \forall p \in (R \cap [r+1, s]; \\ & \text{and } \ z_{(a_{p,v_{p,(r,\rho),(s,\sigma),t}}),(a_{q,v_{q,(r,\rho),(s,\sigma),t}}),(a_{s,\sigma})} > 0, \ \forall \ (p, q) \\ & \in (R \cap [r, s-1])^2 \text{ such that } q > p \ \Big\} \end{aligned} \qquad (2.19)$$

as a "path in (*y*, *z*) from (r, ρ) to (s, σ)." Hence, for convenience, a *path in* (**y**, **z**) *from* (*r*,*ρ*) *to* (*s*, *σ*), $U_{(r,\rho),(s,\sigma),t}(y, z)$, can be alternatively represented as an ordered set of city indices,

$$\begin{aligned} P_{(r,\rho),(s,\sigma),t}(\mathbf{y}, \mathbf{z}) = \langle & i_{r,v_{r,(r,\rho),(s,\sigma),t}}, \\ & i_{r+1,v_{r+1,(r,\rho),(s,\sigma),t}}, \cdots, i_{s+1,v_{s+1,(r,\rho),(s,\sigma),t}} \rangle \end{aligned} \qquad (2.20)$$

where:

$v_{r,(r,\rho),(s,\sigma),t} = \rho$, $v_{s,(r,\rho),(s,\sigma),t} = \sigma$,

$i_{r+1,v_{r+1,(r,\rho),(s,\sigma),t}} = j_{r,\rho}$,

$i_{s+1,v_{s+1,(r,\rho),(s,\sigma),t}} = j_{s,\sigma}$,

$(i_{p,v_{p,(r,\rho),(s,\sigma),t}}, p, i_{p+1,v_{p+1,(r,\rho),(s,\sigma),t}}) \in X_p(y, z)$,

$\forall p \in (R \cap [r, s])$; and

$i_{p,v_{p,(r,\rho),(s,\sigma),t}} = j_{p-1,v_{p-1,(r,\rho),(s,\sigma),t}}$,

$\forall p \in (R \cap [r+1, s]$.

Finally, we denote the set of all *paths in* (**y**, **z**) *from* (*r*,*ρ*) *to* (*s*, *σ*) as $Q_{(r,\rho),(s,\sigma)}(\mathbf{y}, \mathbf{z})$, and associate to it the index set $\Psi_{(r,\rho),(s,\sigma)}(\mathbf{y}, \mathbf{z}) \equiv \{1, 2, \cdots, \varphi_{(r,\rho),(s,\sigma)}(\mathbf{y}, \mathbf{z})\}$, where $\varphi_{(r,\rho),(s,\sigma)}(\mathbf{y}, \mathbf{z})$ is the cardinality of $Q_{(r,\rho),(s,\sigma)}(\mathbf{y}, \mathbf{z})$.

We have the following.

**Theorem 2**

Let $(\mathbf{y}, \mathbf{z}) = ( y_{irjkst}, z_{upvirjkst} )$ be a feasible solution to *Problem LP*. For $(r, s) \in R^2$ $(s \geq r+2)$, $\rho \in N_r(\mathbf{y}, \mathbf{z})$, and $\sigma \in N_s(\mathbf{y}, \mathbf{z})$, if $y_{i_{r,\rho}, r, j_{r,\rho}, i_{s,\sigma}, s, j_{s,\sigma}} > 0$, then we must have:

i) $Q_{(r,\rho),(s,\sigma)}(y, z) \neq \varnothing$; and

ii) $\forall\ g \in (R \cap [r+1, s-1])$ and $\gamma \in N_g(\mathbf{y}, \mathbf{z})$: $( z_{i_{r,\rho}, r, j_{r,\rho},\ i_{g,\gamma}, g, j_{g,\gamma},\ i_{s,\sigma}, s, j_{s,\sigma}} > 0 ) \Rightarrow \exists\ ( \iota \in \Psi_{(r,\rho),(s,\sigma)}(\mathbf{y}, \mathbf{z}) \ni : ( i_{g,\gamma}, j_{g,\gamma} ) \in ( P_{(r,\rho),(s,\sigma),\iota}(\mathbf{y}, \mathbf{z}))^2 )$.

*Proof*:

First, (i) we will show that the theorem holds for all $(r, s) \in R^2$ such that $s = r+2$. Then, (ii) we will show that if the theorem holds for all $(r, s) \in R^2$ such that $s \in [r+2, r+\omega]$ for some integer $\omega \geq 2$, then the theorem must hold for all $(r, s) \in R^2$ such that $s = r+\omega+1$ (if there exists such a pair).

i) Because of constraints 2.16, constraints 2.10 for any $(r, s) \in R^2$ such that $s = r+2$ can be written as:

$$y_{i,r,j,\ u,r+2,v} - z_{i,r,j,\ j,r+1,u,\ u,r+2,v} = 0;$$
$$i \in M;\ j \in M \setminus \{ i \};\ u \in M \setminus \{i, j\};$$
$$v \in M \setminus \{i, j, u\} \qquad (2.21)$$

It follows from 2.21 that for $\sigma \in N_{r+2}(\mathbf{y}, \mathbf{z})$,

$$( y_{i_{r,\rho}, r, j_{r,\rho},\ i_{r+2,\sigma}, r+2, j_{r+2,\sigma}} > 0 ) \Leftrightarrow$$
$$( z_{i_{r,\rho}, r, j_{r,\rho},\ j_{r,\rho}, r+1, i_{r+2,\sigma},\ i_{r+2,\sigma}, r+2, j_{r+2,\sigma}} > 0 ) \qquad (2.22)$$

Hence, for $\sigma \in N_{r+2}(\mathbf{y}, \mathbf{z})$ such that $y_{i_{r,\rho}, r, j_{r,\rho}, i_{r+2,\sigma}, r+2, j_{r+2,\sigma}} > 0$, we have:

$\varphi_{(r,\rho),(r+2,\sigma)}(\mathbf{y}, \mathbf{z}) = 1$, so that:

$Q_{(r,\rho),(r+2,\sigma)}(\mathbf{y}, \mathbf{z}) = \{ P_{(r,\rho),(r+2,\sigma),1}(\mathbf{y}, \mathbf{z}) \}$,

where:

$$P_{(r,\rho),(r+2,\sigma),1}(\mathbf{y}, \mathbf{z}) =$$
$$= \langle i_{r,\rho}, j_{r,\rho}, i_{r+2,\sigma}, j_{r+2,\sigma} \rangle \qquad (2.23)$$

Hence, the theorem holds for all $(r, s) \in R^2$ such that $s = r+2$.

ii) Suppose the theorem holds for all $(r, s) \in R^2$ such that $r+2 \leq s \leq r+\omega$ for some integer $\omega \geq 2$. If $\omega$ is such that there does not exist $(r, t) \in R^2$ with $t = r+\omega+1$, then the theorem is proven. Hence, assume there exist some $(r, t) \in R^2$ such that $t = r+\omega+1$. Consider one such $(r, t)$ pair, and $\tau \in N_t(\mathbf{y}, \mathbf{z})$ such that:

$$y_{i_{r,\rho}, r, j_{r,\rho},\ i_{t,\tau}, t, j_{t,\tau}} > 0 \qquad (2.24)$$

Then, the combination of constraints 2.9, 2.11, 2.16, and condition 2.24 implies that there must exist a set:

$$C_{(r,\rho),(t,\tau)}(\mathbf{y}, \mathbf{z}) \equiv \{ \alpha \in N_{r+1}(\mathbf{y}, \mathbf{z}) \mid$$
$$z_{i_{r,\rho}, r, j_{r,\rho},\ j_{r,\rho}, r+1, j_{r+1,\alpha},\ i_{t,\tau}, t, j_{t,\tau}} > 0 \} \qquad (2.25)$$

such that:

$$y_{i_{r,\rho}, r, j_{r,\rho},\ i_{t,\tau}, t, j_{t,\tau}} = \sum_{\alpha \in C_{(r,\rho),(t,\tau)}(\mathbf{y}, \mathbf{z})} z_{i_{r,\rho}, r, j_{r,\rho},\ j_{r,\rho}, r+1, j_{r+1,\alpha},\ i_{t,\tau}, t, j_{t,\tau}} \qquad (2.26)$$

($C_{(r,\rho),(t,\tau)}(\mathbf{y}, \mathbf{z})$ is the index set of the arcs at stage $r+1$ along which flow from arc $( i_{r,\rho}, r, j_{r,\rho} )$ propagates onto arc $( i_{t,\tau}, t, j_{t,\tau} )$).

By constraints 2.10, expression 2.25 implies:

$$y_{j_{r,\rho}, r+1, j_{r+1,\alpha},\ i_{t,\tau}, t, j_{t,\tau}} > 0$$
$$\forall\ \alpha \in C_{(r,\rho),(t,\tau)}(\mathbf{y}, \mathbf{z}) \qquad (2.27)$$

Hence, by assumption, the theorem holds for $t$, $\tau$, $r+1$, and each $\alpha \in C_{(r,\rho),(t,\tau)}(\mathbf{y}, \mathbf{z})$. Combining this with 2.26, the connectivity requirement constraints 2.8 - 2.11, and the *visit* requirements constraints 2.15, we must have that for all $h \in (R \cap [r+2, t-1])$ and $\mu \in N_h(\mathbf{y}, \mathbf{z})$:

$$\{ z_{i_{r,\rho}, r, j_{r,\rho},\ i_{h,\mu}, h, j_{h,\mu},\ i_{t,\tau}, t, j_{t,\tau}} > 0 \} \Rightarrow$$
$$\exists\ \{ \alpha \in C_{(r,\rho),(t,\tau)}(\mathbf{y}, \mathbf{z}) \text{ and}$$
$$k \in \Psi_{(r+1,\alpha),(t,\tau)}(\mathbf{y}, \mathbf{z}) \ni : ( i_{h,\mu}, j_{h,\mu} ) \in$$
$$( P_{(r+1,\alpha),(t,\tau),k}(\mathbf{y}, \mathbf{z}))^2 \ \}. \qquad (2.28)$$

Condition 2.28 combined with constraints 2.11 – 2.14, and 2.16, imply that:

$$\exists \{ J_{(r+1,\alpha),(t,\tau)}(\mathbf{y}, \mathbf{z}) \subseteq \Psi_{(r+1,\alpha),(t,\tau)}(\mathbf{y}, \mathbf{z}) \} \ni:$$
$$\{ z_{(a_{r,\rho}),(a_{p,v_{p,(r+1,\alpha),(t,\tau),\beta}}),(a_{t,\tau})} > 0 \quad \forall\ (\alpha, \beta) \in$$
$$( C_{(r,\rho),(t,\tau)}(\mathbf{y}, \mathbf{z}),\ J_{(r+1,\alpha),(t,\tau)}(\mathbf{y}, \mathbf{z}) ), \text{ and}$$
$$p \in (R \cap [r+1, t-1]) \} \qquad (2.29)$$

($J_{(r+1,\alpha),(t,\tau)}(\mathbf{y}, \mathbf{z})$ is the index set of the *paths in* $(\mathbf{y}, \mathbf{z})$ *from* $(r+1, \alpha)$ *to* $(t, \tau)$ along which flow from arc $( i_{r,\rho}, r, j_{r,\rho} )$ propagates onto arc $( i_{t,\tau}, t, j_{t,\tau} )$).

Now, for $(\alpha, \beta) \in ( C_{(r,\rho),(t,\tau)}(\mathbf{y}, \mathbf{z}), J_{(r+1,\alpha),(t,\tau)}(\mathbf{y}, \mathbf{z}) )$, let:

$$T_{(r,\rho),(\alpha,\beta)(t,\tau)}(\mathbf{y}, \mathbf{z}) \equiv$$
$$\equiv \{ i_{r,\rho} \} \cup P_{(r+1,\alpha),(t,\tau),\beta}(\mathbf{y}, \mathbf{z}), \qquad (2.30)$$

(where $i_{r,\rho}$ is added to $P_{(r+1,\alpha),(t,\tau),\beta}(\mathbf{y},\mathbf{z})$ in such a way that it occupies the first position in $T_{(r,\rho),(\alpha,\beta)(t,\tau)}(\mathbf{y},\mathbf{z})$ ).

It is easy to verify that $T_{(r,\rho),(\alpha,\beta)(t,\tau)}(\mathbf{y},\mathbf{z})$ is a *path in* $(\mathbf{y},\mathbf{z})$ *from* $(r,\rho)$ *to* $(t,\tau)$. Hence, we have $Q_{(r,\rho),(t,\tau)}(\mathbf{y},\mathbf{z}) \neq \varnothing$. Moreover, it follows directly from 2.28 above that condition ii) of the theorem must hold for r, ρ, t, and τ.

Q.E.D.

**Theorem 3**

Let $(\mathbf{y},\mathbf{z}) = (y_{irjkst}, z_{upvirjkst})$ be a feasible solution to *Problem LP*. Let $(r, s) \in R^2$, $s \geq r+2$; $\rho \in N_r(\mathbf{y},\mathbf{z})$; and $\sigma \in N_s(\mathbf{y},\mathbf{z})$ be such that $y_{i_{r,\rho},r,j_{r,\rho},i_{s,\sigma},s,j_{s,\sigma}} > 0$. Then, we must have:

i) $Q_{(r,\rho),(s,\sigma)}(\mathbf{y},\mathbf{z}) \neq \varnothing$;

Furthermore, for each $\ell \in \Psi_{(r,\rho),(s,\sigma)}(\mathbf{y},\mathbf{z})$ we must have:

ii) $i_{q,v_{q,(r,\rho),(s,\sigma),\ell}} = j_{q-1,v_{q-1,(r,\rho),(s,\sigma),\ell}}$ for $q \in R$; $r+1 \leq q \leq s$;

iii) $z_{(a_{p,v_{p,(r,\rho),(s,\sigma),\ell}}),(a_{q,v_{q,(r,\rho),(s,\sigma),\ell}}),(a_{s,\sigma})} > 0$ $\forall\ (p,q) \in (R \cap [r,s])^2$, $r \leq p < q \leq s-1$;

iv) $i_{p,v_{p,(r,\rho),(s,\sigma),\ell}} \neq i_{q,v_{q,(r,\rho),(s,\sigma),\ell}}$ $\forall\ (p,q) \in (s \cap [r,s+1])^2 \ni:\ p \neq q$.

*Proof*:

Conditions i) –iii) follow directly from definitions and Theorem 2. Condition iv) follows from the combination of condition iii) and the *visit* restrictions constraints 2.16.

Q.E.D.

Hence, every *path in* $(\mathbf{y},\mathbf{z})$ *from* $(1, \bullet)$ *to* $(n-2, \bullet)$ corresponds to a *c.a.s.s. path* of Graph G (and therefore, to a TSP tour). Hence, for convenience, we refer to each $P_{(1,\rho),(n-2,\sigma),k}(y,z)$ simply as a "TSP tour in $(y,z)$," and denote it by $T_{\rho,\sigma,k}(\mathbf{y},\mathbf{z})$. To a *TSP tour in* $(\mathbf{y},\mathbf{z})$, $T_{\rho,\sigma,k}(y,z)$, we attach a "flow value" $\lambda_{\rho,\sigma,k}(y,z)$ defined as:

$$\lambda_{\rho,\sigma,k}(\mathbf{y},\mathbf{z}) \equiv \min_{p \in (R \cap [2,n-3])} \left\{ z_{(a_{1,\rho}),(a_{p,v_{p,(1,\rho),(n-2,\sigma),k}}),(a_{n-2,\sigma})} \right\} \quad (2.31)$$

Let $\Pi(\mathbf{y},\mathbf{z})$ denote the set of all the *TSP tours in* $(\mathbf{y},\mathbf{z})$. Associate to $\Pi(\mathbf{y},\mathbf{z})$ the index set

$$\pi(\mathbf{y},\mathbf{z}) \equiv \{1, 2, \ldots, m(\mathbf{y},\mathbf{z})\},$$

where:

$$m(\mathbf{y},\mathbf{z}) \equiv \sum_{\rho \in N_1(\mathbf{y},\mathbf{z})} \sum_{\sigma \in N_{n-2}(\mathbf{y},\mathbf{z})} \left(\varphi_{(1,\rho),(n-2,\sigma)}(\mathbf{y},\mathbf{z})\right).$$

Rewrite $\Pi(\mathbf{y},\mathbf{z})$ as:

$$\Pi(\mathbf{y},\mathbf{z}) = \{T_{\alpha_p,\beta_p,\kappa_p}(\mathbf{y},\mathbf{z})\ ;\ p = 1, 2, \ldots, m(\mathbf{y},\mathbf{z})\},$$

and denote the arc set associated with $T_{\alpha_p,\beta_p,\kappa_p}(\mathbf{y},\mathbf{z}) \in \Pi(\mathbf{y},\mathbf{z})$, as:

$$\mathbf{a}_p(\mathbf{y},\mathbf{z}) \equiv \{a_{q,v_{q,(1,\alpha_p),(n-2,\beta_p),\kappa_p}}\ ;\ q \in [1, n-2]\}.$$

We have the following:

**Theorem 4**

Let $(\mathbf{y},\mathbf{z}) = (y_{irjkst}, z_{irjupvkst})$ be a feasible solution to *Problem LP*. Then, the following statements are true:

i) $y_{(a_{r,\rho}),(a_{r,\rho})} = \displaystyle\sum_{p \in [1,m(\mathbf{y},\mathbf{z})] \,\|\, a_{r,\rho} \in \mathbf{a}_p(\mathbf{y},\mathbf{z})} \lambda_{\rho_p,\sigma_p,k_p}(\mathbf{y},\mathbf{z})$

$\forall\ (r,\rho) \in (R,\ N_r(\mathbf{y},\mathbf{z}))$;

ii) $y_{(a_{r,\rho}),(a_{s,\sigma})} = \displaystyle\sum_{p \in [1,m(\mathbf{y},\mathbf{z})] \,\|\, (a_{r,\rho},a_{s,\sigma}) \in (\mathbf{a}_p(\mathbf{y},\mathbf{z}))^2} \lambda_{\rho_p,\sigma_p,k_p}(\mathbf{y},\mathbf{z})$

$\forall\ (r,s) \in R^2$, $(\rho,\sigma) \in (N_s(\mathbf{y},\mathbf{z}), N_s(\mathbf{y},\mathbf{z}))$;

iii) $z_{(a_{r,\rho}),(a_{s,\sigma}),(a_{t,\tau})} = \displaystyle\sum_{p \in [1,m(\mathbf{y},\mathbf{z})] \,\|\, (a_{r,\rho},a_{s,\sigma},a_{t,\tau}) \in (\mathbf{a}_p(\mathbf{y},\mathbf{z}))^3} \lambda_{\rho_p,\sigma_p,k_p}(\mathbf{y},\mathbf{z})$

$\forall\ (r,s,t) \in R^3$,
$(\rho,\sigma,\tau) \in (N_s(\mathbf{y},\mathbf{z}), N_s(\mathbf{y},\mathbf{z}), N_t(\mathbf{y},\mathbf{z}))$.

*Proof*:

In the following discussion $m(\mathbf{y},\mathbf{z})$ and the $\mathbf{a}_p(\mathbf{y},\mathbf{z})$ ($p \in [1, m(\mathbf{y},\mathbf{z})]$ will be written simply as $m$ and $\mathbf{a}_p$, respectively, for convenience.

From constraints 2.7-2.10 and Theorem 3, we must have:

$$\bigcup_{p \in \pi(\mathbf{y},\mathbf{z})} (\mathbf{a}_p) = \Lambda(\mathbf{y},\mathbf{z}) \quad (2.32)$$

Also, because of constraints 2.16 and the connectivity requirements 2.11, arcs originating at the same stage of Graph G(y, z) must belong to distinct *TSP tours in* $(y,z)$. Note also that a given *TSP tour in* $(\mathbf{y},\mathbf{z})$ cannot be represented as a convex combination of other *TSP tours in* $(\mathbf{y},\mathbf{z})$. Hence, the flows along distinct *TSP tours in* $(\mathbf{y},\mathbf{z})$ must be additive at any given stage of Graph G(y, z).

We will now consider Conditions i) – iii) in turn.

***Condition i).*** Constraints 2.11 combined with the additivity of the flow amounts discussed above imply that we must have:

$$y_{(a_{1,\rho}),(a_{1,\rho})} = \sum_{\sigma \in N_s(\mathbf{y},\mathbf{z})} \; \sum_{p \in \pi(\mathbf{y},\mathbf{z}) \mid \alpha_p = \rho;\, a_{s,\sigma} \in \mathbf{a}_p} (\lambda_{\alpha_p,\beta_p,\kappa_p}(\mathbf{y},\mathbf{z}))$$

$\forall\, \rho \in N_1(\mathbf{y},\mathbf{z})$; and $s \in R\backslash\{1\}$ (2.33)

From Lemma 1-i), we must also have:

$$y_{(a_{1,\rho}),(a_{1,\rho})} = \sum_{\sigma \in N_s(\mathbf{y},\mathbf{z})} y_{(a_{1,\rho}),(a_{s,\sigma})},$$

$\forall\, s \in R\backslash\{1\}$ (2.34)

Combining 2.33 with 2.34 and re-arranging gives:

$$\sum_{\sigma \in N_s(\mathbf{y},\mathbf{z})} \Bigg( y_{(a_{1,\rho}),(a_{s,\sigma})} + - \sum_{p \in \pi(\mathbf{y},\mathbf{z}) \mid \alpha_p = \rho;\, (a_{s,\sigma}) \in \mathbf{a}_p} (\lambda_{\alpha_p,\beta_p,\kappa_p}(\mathbf{y},\mathbf{z})) \Bigg) = 0$$

$\forall\, \rho \in N_1(\mathbf{y}, \mathbf{z})$, and $s \in R\backslash\{1\}$ (2.35)

From the additivity of the flows along distinct *TSP tours in (y, z)* at any given stage discussed above, we must also have:

$$y_{(a_{1,\rho}),(a_{s,\sigma})} \geq \sum_{p \in \pi(\mathbf{y},\mathbf{z}) \mid \alpha_p = \rho;\, a_{s,\sigma} \in \mathbf{a}_p} (\lambda_{\alpha_p,\beta_p,\kappa_p}(\mathbf{y},\mathbf{z}));$$

$\forall\, \rho \in N_1(\mathbf{y},\mathbf{z})$, $s \in R\backslash\{1\}$, and $\sigma \in N_s(\mathbf{y},\mathbf{z})$ (2.36)

Combining 2.36 and 2.35 gives:

$$y_{(a_{1,\rho}),(a_{s,\sigma})} = \sum_{p \in \pi(\mathbf{y},\mathbf{z}) \mid \alpha_p = \rho;\, a_{s,\sigma} \in \mathbf{a}_p} (\lambda_{\alpha_p,\beta_p,\kappa_p}(\mathbf{y},\mathbf{z}))$$

Condition i) follows directly from this, relations 2.33, and constraints 2.7.

***Condition ii.*** From Theorem 3, we have:

$\forall\, (r, s) \in R^2, r < s,\ \rho \in N_r(\mathbf{y},\mathbf{z})$, and $\sigma \in N_s(\mathbf{y},\mathbf{z})$,

$$( y_{(a_{r,\rho}),(a_{s,\sigma})} > 0 ) \Leftrightarrow \exists\, ( p \in \pi(\mathbf{y},\mathbf{z}) \ni: (a_{r,\rho}, a_{s,\sigma}) \in \mathbf{a}_p^2 ). \quad (2.37)$$

Combining 2.37 with Condition 2.32, we must have:

$$y_{(a_{r,\rho}),(a_{r,\rho})} = \sum_{\sigma \in N_s(\mathbf{y},\mathbf{z})} \; \sum_{p \in \pi(y,z) \mid (a_{r,\rho}, a_{s,\sigma}) \in \mathbf{a}_p^2} (\lambda_{\alpha_p,\beta_p,\kappa_p}(\mathbf{y},\mathbf{z}))$$

$\forall\, (r, s) \in R^2, s > r$; and $\rho \in N_r(\mathbf{y},\mathbf{z})$ (2.38)

Also, from Lemma 1-i), we must have:

$$y_{(a_{r,\rho}),(a_{r,\rho})} = \sum_{\sigma \in N_s(\mathbf{y},\mathbf{z})} y_{(a_{r,\rho}),(a_{s,\sigma})},$$

$\forall\, (r, s) \in R^2, s > r$; and $\rho \in N_r(\mathbf{y}, \mathbf{z})$ (2.39)

Combining 2.38 with 2.39 and re-arranging gives:

$$\sum_{\sigma \in N_s(y,z)} \Bigg( y_{(a_{r,\rho}),(a_{s,\sigma})} + - \sum_{p \in \pi(\mathbf{y},\mathbf{z}) \mid (a_{r,\rho}, a_{s,\sigma}) \in \mathbf{a}_p^2} (\lambda_{\alpha_p,\beta_p,\kappa_p}(\mathbf{y}, \mathbf{z})) \Bigg) = 0$$

$\forall\, (r, s) \in R^2, s > r$; and $\rho \in N_r(y, z)$ (2.40)

From 2.37 and the additivity of the flows along distinct *TSP tours in (***y***,* **z***)* at any given stage discussed above, we must also have:

$$y_{(a_{r,\rho}),(a_{s,\sigma})} \geq \sum_{p \in \pi(\mathbf{y},\mathbf{z}) \mid (a_{r,\rho}, a_{s,\sigma}) \in \mathbf{a}_p^2} (\lambda_{\alpha_p,\beta_p,\kappa_p}(\mathbf{y},\mathbf{z}))$$

$\forall\, (r, s) \in R^2, s > r$; and

$(\rho, \sigma) \in ( N_r(\mathbf{y},\mathbf{z}), N_s(\mathbf{y},\mathbf{z}) )$ (2.41)

Condition ii) follows directly from the combination of 2.40 and 2.41.

***Condition iii).*** The proof for Condition iii) is similar to that of Condition ii) (although it uses Lemma 1-ii) instead of Lemma 1-i)) and is therefore omitted.

Q.E.D.

Hence, any given feasible solution to Problem LP, (*y*, *z*), must be a convex combination of the *feasible solutions corresponding to the TSP tours in (***y***,* **z***)* with weights equal to the associated *flow values*, respectively.

**Theorem 5**

The following statements are true of basic feasible solutions (BFS) of *Problem LP* and TSP tours:

1) Every BFS of *Problem LP* corresponds to a TSP tour;
2) Every TSP tour corresponds to a BFS of *Problem LP*;
3) The mapping of BFS's of *Problem LP* onto TSP tours is surjective.

*Proof*:

1) Correspondence of a BFS of *Problem LP* to a TSP tour follows from the fact that every TSP tour corresponds to a feasible solution to

*Problem LP* (Theorem 1), the fact that every feasible solution to *Problem LP* corresponds to a convex combination of TSP tours (Theorem 4), and the fact that a BFS cannot be a convex combination of other feasible solutions.

2) Correspondence of a TSP tour to a BFS of *Problem LP* follows from Theorem 1, Theorem 4, and the fact that a given TSP tour cannot be represented as a convex combination of other TSP tours.

3) It easy to verify that the number of non-zero components of the *feasible solution corresponding to a given TSP tour* is less than $n^3$, and that the number of constraints of *Problem LP* exceeds $n^3$. Hence, Statement 1) of the theorem implies that there must be basic variables that are equal to zero in any BFS of *Problem LP*. The surjective nature of the "BFS's-to-TSP tours" mapping follows from this and the fact that BFS's of *Problem LP* that have the same set of positive variables in common correspond to the same TSP tour.

Q.E.D.

**Corollary 1**

Let *Conv((•))* denote the convex hull of the feasible set of *Problem (•)*. Then, we have:

*Conv(LP) = Conv(IP).*

**Corollary 2**

*Problem LP* and *Problem IP* (and therefore, *Problem TSP*) are equivalent.

**Theorem 6**

Computational complexity classes *P* and *NP* are equal.

*Proof:*

First, note that *Problem LP* has $O(n^9)$ variables and $O(n^8)$ constraints. Hence, it can be explicitly stated in polynomial time. The theorem follows from this, Corollary 2, the NP-Completeness of the TSP decision problem (see Garey and Johnson [1979], or Nemhauser and Wolsey [1988], among others), and the fact that an explicitly-stated instance of *Problem LP* can be solved in polynomial-time (see Katchiyan [1979], or Karmarkar [1984]).

Q.E.D.

## 3 Numerical Implementation

In implementing the model, we replaced constraints 2.18 with simple non-negativity constraints on the $y_{irjkst}$ and $z_{irjupvkst}$ variables (since the upper bounds in those constraints are redundant according to Theorem 4). Also, we did not explicitly consider constraints 2.16 and the variables they restrict to zero, and accordingly re-wrote/expanded the other constraints of the model.

We used the simplex method implementation of the *OSL* optimization package (IBM) to solve a set of randomly-generated 7-city problems. The travel costs in these randomly-generated problems were taken as uniform integer numbers between 1 and 300. Three of these problems had symmetric costs. The other three randomly-generated problems had asymmetric costs. We also solved an additional set of 7-city problems we refer to as "extreme-symmetry" problems. These "extreme-symmetry" problems are labeled "*xtsp71*," "*xtsp72*," and "*xtsp73*," respectively. In *Problem xtsp71*, all travel costs, $t_{ij}$, are equal to (-1), except for $t_{12}$ and $t_{21}$ which are equal to 1, respectively. In *Problem xtsp72*, all travel costs, $t_{ij}$, are equal to 1, except for $t_{12}$ and $t_{21}$ which are equal to (-100), respectively. Finally, in *Problem xtsp73*, all travel costs, $t_{ij}$, are equal to 0, except for $t_{12}$ and $t_{21}$ which are equal to 1, respectively.

We solved both the dual and primal forms of each of the test problems described above, respectively. The computational results are summarized in Table 3.1 (More details can be found in Diaby [2007]).

Using the dual forms, the averages of the numbers of iterations were 475.0, 1,752.7, and 3,880.5 for the asymmetric, symmetric, and "extreme-symmetry" problems, respectively. The corresponding average computational times were 0.1617, 1.3493, and 9.0785 CPU seconds of Sony VAIO VGN-FE 770G notebook computer (1.8 GHz Intel Core 2 Duo Processor) time, respectively.

For the primal forms, the average number of iterations was 2,203.0, 3,542.0, and 3,315.7 for the asymmetric, symmetric, and "extreme-symmetry" problems, respectively. The corresponding average computational times were 2.8910, 6.5157, and 5.4900 CPU seconds, respectively. The average number of TSP tours examined in the simplex procedure was 1.0, 1.3, and 1.0 for the asymmetric, symmetric, and "extreme-symmetry" problems, respectively.

Overall, we believe our computational experience provided the empirical validation of our theoretical developments in section 2 of this paper that we expected. The dual forms outperformed the primal forms in general. However, the primal form appears to hold some promise with respect to future developments aimed at solving large-sized problems because of the small number of TSP tours that are examined when the primal form is used.

## 4 Conclusions

We have presented a first polynomial-sized linear programming formulation of the TSP. Our approach can be used to formulate general integer programming problems as linear programs, since the general integer programming problem is *polynomially transformable* to a Hamiltonian Path problem (see Johnson and Papadimitriou [1985, pp. 61-74]). Note however, that the Hamiltonian Path problem resulting from the transformation involved is very-large-scale. Hence, we believe a key issue at this point is the question of whether the suggested modeling approach can be developed into a more general, unified framework that would extend in a more natural way to other *NP-Complete* problems (see Garey and Johnson [1979], or Nemhauser and Wolsey [1988], among others).

| Problem | | Dual Form | | Primal Form | | |
|---|---|---|---|---|---|---|
| Name[1] | Value | # of Iter.[2] | CPU Sec.[3] | # of Tours[4] | # of Iter.[2] | CPU Sec.[3] |
| atsp71 | 414 | 643 | 0.235 | 1 | 3,541 | 6.797 |
| atsp72 | 468 | 471 | 0.141 | 1 | 1,797 | 1.563 |
| atsp73 | 354 | 311 | 0.109 | 1 | 1,271 | 0.313 |
| **Ave.** | — | **475.0** | **0.1617** | **1.0** | **2,203.0** | **2.8910** |
| stsp71 | 503 | 1,954 | 1.938 | 1 | 3,476 | 6.657 |
| stsp72 | 531 | 1,790 | 1.188 | 2 | 3,572 | 6.093 |
| stsp73 | 637 | 1,514 | 0.922 | 1 | 3,578 | 6.797 |
| **Ave.** | — | **1,752.7** | **1.3493** | **1.3** | **3,542.0** | **6.5157** |
| xtsp71 | (-7) | 4,361 | 12.141 | 1 | 3,307 | 5.094 |
| xtsp72 | (-94) | 3,201 | 5.735 | 1 | 3,240 | 5.360 |
| xtsp73 | 0 | 3,400 | 6.016 | 1 | 3,400 | 6.016 |
| **Ave.** | — | **3,880.5** | **9.0785** | **1** | **3,315.7** | **5.4900** |

1: "atsp··": asymmetric costs; "stsp··": symmetric costs; "xtsp··": "extreme symmetry" problem

2: Number of simplex iterations

3: Sony VAIO VGN-FE 770G notebook computer (1.8 GHz Intel Core 2 Duo Processor)

4. Number of TSP tours examined in the simplex procedure

Table 3.1: *Summary of the Computational Results*